\documentclass[12pt]{article}

\usepackage{graphicx}
\usepackage{amssymb,amsmath}
\usepackage{url}
\usepackage{natbib}
\usepackage[format=hang,font=small,labelfont=bf]{caption} 
\usepackage[none]{hyphenat} 
\usepackage{hyperref}

\bibpunct{(}{)}{;}{a}{}{,} 

\begin{document}

\title{Estimating robustness of the tileShuffle method with repeated probes}

\author{Sigrun Helga Lund\,$^{1,3}$, Asgeir Sigurdsson\,$^{2}$,
Sigurjon Axel Gudjonsson\,$^2$, \\
Julius Gudmundsson\,$^2$,  Daniel Fannar Gudbjartsson\,$^2$, \\
Thorunn Rafnar\,$^2$, Kari Stefansson\,$^2$ and Gunnar Stefansson$^{1}$}

\footnotetext[1]{University of Iceland, Science Institute, Dunhaga 3, 107 Reykjavik, Iceland}
\footnotetext[2]{deCODE Genetics, Sturlugata 8, 101 Reykjavik, Iceland.}
\footnotetext[3]{to whom correspondence should be addressed}




\maketitle

\begin{abstract}
\textbf{Motivation:}

In this paper the TileShuffle method is evaluated as a search method
for candidate lncRNAs at 8q24.2. The method is run on three
microarrays. Microarrays which all contained the same sample and
repeated copies of tiled probes.  This allows the coherence of the
selection method within and between microarrays to be estimated by
Monte Carlo simulations on the repeated probes.

\textbf{Results:} 
The results show poor consistency in areas selected between arrays containing identical samples. A crude application
of the method can result in majority of the region to be selected,
resulting in a need for further restrictions on the selection.
Restrictions based on ranking internal tileShuffle test statistics do not increase precision.
As the tileShuffle method has been shown to have higher precision than the
MAS and TAS software, one can conclude that methods giving unreliable
results are in common use.  

\textbf{Availability:} 
The data discussed in this publication have been deposited in NCBI's
Gene Expression Omnibus and are accessible through GEO Series
accession number GSE45934.

\textbf{Contact:} sigrunhelga@gmail.com
\end{abstract}


\section{Introduction}

Gene expression is the process by which information from a gene is
used in the synthesis of a functional gene product. These products are
often proteins, but there are also non-protein coding genes where
the product is a functional RNA. It has been predicted that more than
30,000 RNA genes are associated with the human genome
\citep{kapranov2012dark}.

Non-protein coding genes and their products can vary considerably in
length. The shortest products, micro RNAs (miRNA), are on average only
22 bp, whereas long non-coding RNA (lncRNAs) are defined as
transcribed RNA molecules longer than 200 nucleotides in length
\citep{sana2012novel}. 

There have been several publications indicating that lncRNAs might
play an important role in cancer development \citep{wang2011long,huarte2010large,tsai2011long,gibb2011human,shore2012noncoding,willard2012regulators}
and a good review of their functional role in human carcinomas is
given in \cite{gibb2011functional}.

lncRNAs are also thought to play a
regulatory role in cancer-associated pathways governing mechanisms
such as cell growth, invasion, and metastasis and have been seen to be
expressed differently in primary and metastatic cancer \citep{tahira2011long}. lncRNAs might thus provide insights into the
mechanisms underlying tumor development.

lncRNAs originate everywhere in the genome, but especially in long
stretches where no protein-coding genes have been
identified {chung2011association}. An example of such area is 8q24, where multiple single nucleotide polymorphisms (SNPs) have been associated 
 with risk of developing prostate cancer \citep{haiman2007mrw,yeager2007gwa,gudmundsson2007gwa,amundadottir2006cva}.
Currently, there are at least
11 databases which record lncRNAs \citep{dinger2009nred,amaral2011lncrnadb,bu2012noncode,risueno2010gatexplorer,gibb2011functional}.
 

Microarrays are frequently used to locate RNA
genes. A microarray contains multiple copies of the same DNA oligonucleotides, known as probes, which are hybridized to a
labeled RNA sample and the array is subsequently washed. Theoretically
this will result in the labeled sample only remaining where
the sample hybridized to probes. The signal intensities at the
corresponding location on the microarray are used as a measure of the
relative abundance of hybridization of each probe.

Typically a probe corresponds to a specific genomic
region. Sometimes the probes overlap, referred to as tiling, and such
arrays are called tiled microrarrays. Tiled
microarrays have been successful in assessing expression of
non-coding RNAs. \citep{johnson2005dark,mockler2005adt,weile2007use}

The ability to accurately detect the true gene-expression signal in
microarrays is affected by several sources of variation\citep{pozhitkov2007oligonucleotide,churchill2002fed}. Further
issues and different biases arise when using tiled microarrays, 
as opposed to other analysis of differential expression \citep{royce2005issues}.
It is therefore important to take technical
variation into account when doing statistical analysis on microarray
data \citep{wu2004mbb,royce2005issues}.

Currently a variety of methods are available to analyse data from tiled microarrays, but as expression levels are generally lower for lncRNA than protein coding genes,\citep{gibb2011functional} conventional methods for differential expression detection may have difficulties detecting them. A good overview of available methods is found in Otto et al.\citep{otto2012detection}, where the tileShuffle method is introduced and shown to have higher precision than the commonly used TAS \citep{kampa2004novel} and MAT \citep{johnson2006model} methods. 

The tileShuffle method identifies transcribed segments in terms of significant differences from the background distribution, using a permutation test statistic, called a window score. All probes within a sliding window have a window score assigned (arithmetic mean trimmed by median or max and min value). Further, probes are subdivided into bins by GC content and processed independently. The significance of a window score is assessed by permuting probes accross the array, but always within the same bin. Empirical p-values are estimated by counting the number of permuted windows with higher score \citep{otto2012detection}.

The aim of this study is to assess the robustness of the tileShuffle method on the expressed regions level. It utilizes a special array-design where every probe is repeated ten times on each tiled array. This enables Monte-Carlo simulations of expression signals which are used to estimate expression on pseudo-arrays, whose differences lie in a variability that is usually neglected in microarray experiments.  Further, a single biological sample was split in three and used on repeated arrays, providing estimates of  another variability that is commonly neglected. The consistency in regions selected across the pseudo-arrays, that should "in theory" give identical results, will be used as a measure of the robustness of the tileShuffle method.

\section{Methods}
The data discussed in this publication are RNA expression data from custom designed Nimblegen microarray experiments where 
the same prostate tissue sample was used on three arrays. 
The data have been deposited in NCBI's Gene
Expression Omnibus \citep{edgar2002gene} and are accessible through GEO Series accession number GSE45934
\footnote{http://www.ncbi.nlm.nih.gov/geo/query/acc.cgi?acc=GSE45934}.

The arrays contained 50 nucleotide probes from 
chr8:127640000-129120000 at locus 8q24, tiled at a 20 base interval.  The whole region was tiled evenly, but probes with blat score greater than 5 \citep{kent2002bbl} or blast score greater than 40 \citep{altschul1990bla} were excluded. That left in total 54236 (out of 74000) probes, each of which was replicated 10 times on the array.

Spatial artifacts in the expression signal were minimized by aggregating the wells of the microarray into ten non-overlapping logical virtual ''containers'', allocating each of the ten replicates af a probe to a different container.

These ten replicate spots for each probe,
evenly spread across the array, permitted Monte-Carlo simulation
of the expression signals. In that way, for each of the three microarrays, 1000 pseudo replicate arrays were produced, with only one repetition of each probe, selected at random. The peudo-arrays were made in triplets, such that the same set of replicates was used to produce pseudo-arrays for all three microarrays within every simulation.

The arrays were normalized by the quantile normalization method \citep{bolstad2003comparison} and consequently analysed with the tileShuffle method under various conditions, detailed later, but always one at a time. The window size, the minimum length of selected areas, was set as 1000 bases, as the aim was to detect relatively long areas. Three GC-content bins were used and the number of permutations was 1000. All statistical analyses were performed in the R statistical package \citep{rpackage} and graphics are generated with the ggplot2 library \citep{ggplot2}.

The tileShuffle method assesses significance on minimal expected transcriptional units rather than on a single probe level. \citep{otto2012detection} Therefore the tiled region was split up into areas of length 100 bases. These areas will be underlying when addressing genomic locations that are expressed. An area will be deemed as expressed if all corresponding 100 bases were within an expressed region. 

\section{Results}
\subsection{Consistency within an array}
As each probe was repeated ten times on every array, the consistency of the method could be estimated by Monte-Carlo simulations on the probe sets. In that way, 1000 pseudo-arrays were produced, each by randomly  selecting one repetition of each probe. All experimental sources of variation of these pseudo-arrays are identical, except the physical location of the probesets within the microarray.  

\begin{figure}[!tpb]
\includegraphics[width=8cm]{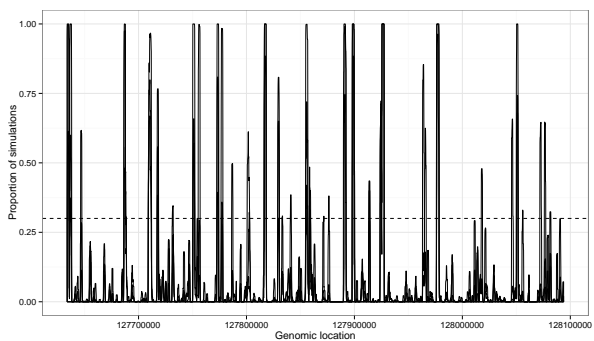}
\caption{A smoothed plot of on how many of the 1000 pseudo-arrays each genomic location was called expressed by the tileShuffle method. This is shown for the first 300.000 bases of the tiled area.}
\label{fig1}
\end{figure}

In order to investigate the effect of this probe-to-probe variation on
calls of expression, the tileShuffle method was run on each of the
1000 pseudo-arrays. Figure \ref{fig1} shows on how many of the
pseudo-arrays each genomic location was "called" "expressed" (p $<$0.05, adjusted for multiplicity). For clarity, the figure shows only the first 300.000 bases of
the tiled area or about one-fifth of the tiled region. Graphs for the
remaining regions were similar.  The figure shows that a great majority of the underlying tiled region is selected on at most 25\% of the pseudo-arrays, whereas a few areas are selected consistently in near all cases.

\subsection{Consistency between arrays}

To compare across arrays, pseudo arrays were simulated in triplicates
so that a single set containing probes from the same physical location
on the microarray was generated for all three arrays at a time. Thus,
within each triplicate of pseudo-arrays, all probes have the same
internal physical location of probes on the original microarrays,
blocking the location effect of probes within a microarray. This emulates real situations where probes are not replicated.

The performance of different methods for selecting subsets of
the areas deemed expressed is compared in Table \ref{Tab:01}. The
three columns show the areas selected on exactly one, two or all three
pseudo-arrays within a triplicate, as a proportion of areas that are
selected on \textbf{some} pseudo-array within the triplicate. Ideally one would like to maximize the proportion of instances where a
location is expressed on either none, or all of the three arrays. The
results are shown by increasing proportion of areas that are selected
on all three pseudo-arrays.

\begin{table}[!t]
\caption{The average proportion of areas that are selected on exactly one, two or all three pseudo-arrays within a triplicate of all areas that are selected on some triplicate. From top-down: 1) Only the 30 areas with the highest window score are selected in every simulation, 2) all the probeset, one replication of each probe, 3) half of the probeset was used with two replicates of each probe, 4) the median score over every 10 probes was calculated a priori and fed to the method, 5) only select areas that are deemed expressed in at least 99\% of the replications 6) all 10 replicates were fed to the method
 \label{Tab:01}}
{\begin{tabular}{llll}\hline
 &One array & Two arrays & Three arrays \\ \hline
thirty highest window scores & 0.58 & 0.19 & 0.22 \\
one replicate of all probes & 0.49 & 0.23 & 0.28 \\ 
two replicates of half of the probes & 0.47 & 0.23 & 0.30 \\ 
median of every ten probes & 0.46 & 0.23 & 0.31 \\
selected in at least 99\% of simulations & 0.52 & 0.16 & 0.32 \\
all ten replicates & 0.23  & 0.14 & 0.63 \\\hline
\end{tabular}}{}
\end{table}

Finally, Supplementary Fig. 1 shows the relationship between the average proportion of the total underlying genomic area that is chosen in each simulations against the number of replicates used in every simulation. The relationship is shown for the proportion that is selected on at least one array, at least two arrays and all three arrays and a number of replicates running from 1 up to 10 replicates of each probe per array.


\section{Discussion}

This paper is based on an experimental setup using three tiled
microarrays containing the same biological sample, each using
ten repetitions of each probe. Monte-Carlo simulations from real data are
used to investigate the robustness of the tileShuffle method when
targeting areas on locus 8q24 that are expressed in prostate cancer.

This study raises several concerns regarding the consistency of areas
selected. First of all the method shows considerable variability
depending on which of the 10 replicates of each probe the method is
applied to. Ideally, every area on Figure \ref{fig1} should be
expressed in either all or none of the Monte-Carlo simulations,
resulting in the proportion being close to 0 or 1. As shown on
the figure, these proportions span the whole spectrum
from zero to one. Most probes which are "called" are only called in fewer than 25\% of all simulations,
 indicating a serious lack of repeatability. A few areas are selected consistently on nearly all pseudo-arrays, but as probes are not repeated 
 in the common situation, this plot is not available and one cannot identify locations that are consistently expressed across pseudo-arrays.

Table \ref{Tab:01} shows poor between-array consistency in
choice of areas, which should ideally be identical. Probes which show consistency across
repetitions within an array (selected in at least 99\% of simulations) do not show more consistency across arrays.

The difference
in results by applying the method on all ten replicates v.s. first
calculating the median of every ten probes and then applying the
method is somewhat counter-intuitive. A few further points should also
be noted: Selecting a fixed number of areas with the highest window
score is an unrobust method. The difference in using 
repeated probes rather then denser tiling is small, although in favour of repeated probes.

Finally, as shown on Supplementary Fig. 1, the average proportion of the underlying genomic region that is selected increases rapidly as the number of replicates of each probe increases. With ten replicates of each probe, almost 60\% of the underlying region is selected on at least one array out of three. This might suggest that  majority of the underlying region is "expressed" by definition and the lack of consistency is in caused by small power.

\section{Conclusion}

This paper shows poor consistency of the tileShuffle method both
between selection of replicates of probes within a microarray and also
between microarrays containing the same sample. As the tileShuffle method
has shown to have higher precision than the MAS and TAS software, one
can conclude that methods giving unreliable results are in common use.

\section{Acknowledgement}

This research project was funded in part by grant 5R01CA129991-02 from the NCI and by 
an FS-Grant from the Icelandic Centre for Research (RANNIS).

\end{document}